\def\dbox#1{\hbox{\vrule  
                        \vbox{\hrule \vskip #1
                             \hbox{\hskip #1
                                 \vbox{\hsize=#1}%
                              \hskip #1}%
                         \vskip #1 \hrule}%
                      \vrule}}
\def\square{\dbox{0.02true in}} 
\documentclass{iopart}
\usepackage{graphicx}
\usepackage{cite}
\usepackage{lscape}
\usepackage{longtable}
\begin{document}



\title{The Quantum Effective Mass Hamilton-Jacobi Problem}

\markboth{ \"O. Ye\c{s}ilta\c{s}}{The Quantum Effective Mass Hamilton-Jacobi Problem}

\author{ \"Ozlem Ye\c{s}ilta\c{s}$^{\dagger}$}
\address{Department of Mathematics and Statistics, Concordia
University,\\ 1455 de Maisonneuve Boulevard West, Montr\'eal,
Qu\'ebec, Canada H3G 1M8. \\Email: yesiltas@gazi.edu.tr\\
$^{\dagger}$Permanent address: Department of Physics, Faculty of Arts and Sciences,
Gazi University, 06500 Ankara, Turkey.}

\begin{abstract}
In this article, the quantum Hamilton- Jacobi theory based on the position dependent mass model is studied. Two effective mass functions having different singularity structures are used to examine the Morse and P\"{o}schl-Teller potentials. The residue method is used to obtain the solutions of the quantum effective mass- Hamilton Jacobi equation. Further, it is shown that the eigenstates of the generalized non-Hermitian Swanson Hamiltonian for Morse and P\"{o}schl-Teller potentials can be obtained by using the Riccati equation without solving a differential equation.
{\small \sl
\\
 \noindent \\
Key words:  position dependent, pseudo-Hermitian, quantum Hamilton Jacobi \\[0.2cm]
}
\end{abstract}
\baselineskip 0.9cm
\newpage
\section{Introduction}
Quantum Hamilton-Jacobi (QHJ) theory has always attracted much attention in both non-relativistic and relativistic quantum mechanics. An interesting work on the quantization of bosonic chiral Schwinger model is studied within HJ context\cite{dimit}. QHJ theory, which is a generalization of  classical Hamilton Jacobi formulation, based on transformation theory \cite{babalar}. A fundamental work for obtaining the eigenfunctions of a quantum system with using the QHJ theory may be found in \cite{bh}. Furthermore, there has been an increased research interest in quantum trajectories in the complex plane \cite{wyatt,wyatt1}. The study of the De Broglie-Bohm approach is formulated in a time dependent domain \cite{john}. The QHJ formalism, first developed by Leacock and Padgett in 1983 \cite{lea}. Exact and quasi-exact solutions of a group of soluble potentials are also studied \cite{ranj} as well as the supersymmetric quantum mechanics and shape invariance approach \cite{asim}. Moreover, Sanz et al showed that the  chemical reactions are analyzed within generalized HJ framework \cite{sanz2}. Some detailed and interesting work on extensions of analytical mechanics to the complex plane can be found in \cite{yang}. In \cite{ozlem}, the spectrum for the class of potentials within QHJ is obtained.

On the other hand, position dependent mass model has been a subject of considerable interest in recent years \cite{bagque,cluster,bastard,fermi,leblond,mazhar,many,roos,liquids,dots} following the work of Von Roos and Levy-Leblond \cite{leblond,roos} that claims to describe some physical phenomena. The importance of this kind of Hamiltonian arises from the motion of electrons and holes in semiconductors that can be described by them. In this model, the mass is taken as the effective mass of the particle, which depends on the material  \cite{leblond}. Von Roos showed that Hamiltonians with position dependent mass are not Galilean invariant within the extension of Bargmann's theorem \cite{roos}. Later, Levy-Leblond showed that Galilean invariance can be used to obtain Hamiltonian of an electron with spatially dependent mass, a model of a heterojunction is proposed in \cite{leblond}.

Recently, a great attention has been devoted to investigate the properties of non-Hermitian Hamiltonians in theoretical physics \cite{others,Bagchitanaka}. A wide class of $\mathcal{PT}$ symmetric, non-Hermitian Hamiltonians provides entirely real spectra \cite{bender}. For recent developments please see \cite{ben1,susy,Jones,znojil}. Mostafazadeh has investigated the larger class of Hamiltonian models known as pseudo-Hermitian, defined by the relation $H^{\dag}=\eta H \eta^{-1}$, for a Hermitian operator $\eta$ \cite{mostafazadeh, mus,mmm}. Various papers appeared on this topic \cite{musumbu,quesne,roy, swanson}. As a non-Hermitian and $\mathcal{PT}$-symmetric Hamiltonian model, some properties of Swanson Hamiltonian are studied in the literature \cite{musumbu,Jones,swanson,others}.

In the present paper, as a first study of the QHJ equation introduced in a mass dependent form that can be a generalization of QHJ equation, exponential and hyperbolic type effective mass models are considered and the transformed equation, which is a Riccati type, is mass dependent. The effective mass-QHJ equation is solved for the effective potentials using a Laurent series. Moreover, a pseudo-Hermitian Hamiltonian model is connected with QHJ equation and two types of effective potentials are generated.
\section{ Preliminary Exploration of Two Approaches}
In this section we shall introduce a brief summary of the QHJ theory and the effective mass problem.
\subsection{Quantum Hamilton Jacobi theory}
In \cite{lea}, the authors considered a particle moving in one dimension under the influence of a potential $U(x)$. Classical transformation equations $p=\frac{\partial W}{\partial x}$, $Q=\frac{\partial W}{\partial P}$ are used in the Schr\"{o}dinger representation which is $\hat{H}=H(\hat{x},\hat{p})=\hat{p}^{2}+U(\hat{x})$ where $\hat{x}$, $\hat{p}$ are linear coordinate and momentum operators. Then, the QHJ equation is stated \cite{lea} as
\begin{equation*}
    \frac{\hbar}{i} \frac{\partial^{2}W(x,E)}{\partial x^{2}}+\left(\frac{\partial W(x,E)}{\partial x}\right)^{2}=E-U(x),
\end{equation*}
where $W(x,E)$ is the generating function. If the quantum momentum function $p$ is defined by $p(x,E)=\partial W(x,E)/\partial x$, then the QHJ equation is defined by
\begin{equation}\label{QHJ}
    p^{2}-i\hbar p^{'}-(E-U(x))=0.
\end{equation}
There can be a transition between the Schr\"{o}dinger and the QHJ equations via the wave-function defined by $\psi(x,E)=e^{\frac{i}{\hbar}W(x,E)}$ \cite{lea}, which satisfies the physical boundary conditions. (\ref{QHJ}) is a nonlinear equation and it has two solutions. But we have to look for a physically acceptable solution. The appropriate boundary condition was proposed as $ p(x,E)\rightarrow_{\hbar\rightarrow 0} p_{cl}(x,E)$ in Leacock and Padgett`s paper; this is only for simple potentials, because  $p_{cl}
(x,E)=\sqrt{E-U(x)}$  may have several branch points. The quantum action variable is defined \cite{lea} by
\begin{equation}
    J(E)=\frac{1}{2\pi}\oint_{C} p(x,E) dx=n\hbar,
\end{equation}
which connects eigenvalue $J$ to the energy $E$. Here, $C$ is the closed contour which encloses the cut of
$p_{cl}(x,E)$ between $x_{1}$ and $x_{2}$. Poles of $p(x,E)$ are enclosed by the contour C, thus $J(E)=n \hbar$ \cite{lea}. As it is seen from quantum action variable and boundary condition that $p(x,E)$ has poles of residue
$-i\hbar$ on the real axis. The details of the model and calculations for obtaining the spectrum will not be given in this paper, however the reader can look at \cite{lea, bh, ranj} for more details.
\subsection{Effective mass problem }
Let us start with the general Hermitian effective mass Hamiltonian which was suggested first by von Roos \cite{roos}
\begin{equation}\label{roos}
    H=\frac{1}{4}(m^{\alpha}(x)\textbf{p}m^{\beta}(x)\textbf{p}m^{\gamma}(x)+
    m^{\gamma}(x)\textbf{p}m^{\beta}(x)\textbf{p}m^{\alpha}(x))+V(x),
\end{equation}
with the constraint $\alpha+\beta+\gamma=-1$ on the parameters, and $m(x)$ is the position dependent effective mass. One can express this Hamiltonian in various ways, depending on the choice of the parameters. Thus, various special cases of the kinetic energy operator can be given \cite{bastard} by :
\begin{eqnarray}
  \hat{T} &=& \frac{1}{4}\left(\frac{1}{m}p^{2}+p^{2}\frac{1}{m}\right) \\
  \hat{T} &=& \frac{1}{2}\left(\textbf{p}\frac{1}{m}\textbf{p}\right) \\
  \hat{T} &=& \frac{1}{2}\left(\frac{1}{\sqrt{m}}p^{2}\frac{1}{\sqrt{m}}\right).
\end{eqnarray}
In this work, we will keep the same general form of the Hamiltonian. Thus, the one-dimensional effective PDM Hamiltonian is given \cite{roos} by
\begin{equation}\label{pdm}
    H_{eff}=-\frac{1}{M(x)}\frac{d^{2}}{dx^{2}}+\frac{M^{'}(x)}{M^{2}(x)}\frac{d}{dx}+V_{eff}(x),
\end{equation}
where
\begin{equation}\label{veff}
    V_{eff}(x)=V(x)+\frac{1}{2}(\beta+1)\frac{M^{''}}{M^{2}}-[\alpha(\alpha+\beta+1)+\beta+1]\frac{(M^{'})^{2}}{M^{3}}
\end{equation}
$\alpha, \beta$ are parameters as written before, and primes stand for the derivatives with
respect to $x$. Note that we use the dimensionless form $M(x)$ of $m(x)=m_{0} M(x)$ and $\hbar=2m_{0}=1$. Now we can introduce the corresponding eigenvalue
equation for (\ref{pdm}) as
\begin{equation}\label{sCH}
    -\frac{1}{M(x)}\frac{\varphi^{''}(x)}{\varphi(x)}+\frac{M^{'}(x)}{M^{2}(x)}\frac{\varphi^{'}(x)}{\varphi(x)}+(V_{eff}(x)-\varepsilon)=0,
\end{equation}
where $\varepsilon$ and $\varphi(x)$ are the eigenvalues and eigenfunctions of the Hamiltonian (\ref{roos}). Our task is now to adapt (\ref{sCH}) to QJH theory.
\section{Quantum Effective Mass- Hamilton Jacobi Model }
Now we make a new transformation as $\tilde{p}(x,\varepsilon)=-i\frac{\varphi^{'}}{\varphi}$ in (\ref{sCH}). Hence we have,
\begin{equation}\label{pdmqhj}
    -i \tilde{p}^{'}(x,\varepsilon)+\tilde{p}^{2}(x,\varepsilon)+i\frac{M^{'}(x)}{M(x)}\tilde{p}(x,\varepsilon)+M(x)[V_{eff}(x)-\varepsilon]=0
\end{equation}
that will be called as the quantum effective mass(QEM)-Hamilton Jacobi(HJ) equation. We shall use $\tilde{p}$ for quantum effective mass momentum function. It is obvious that the QEM-HJ equation can be transformed into the QHJ equation by using
\begin{equation}\label{schrö}
    \psi(x)=\frac{1}{\sqrt{M(x)}} \exp\left(i\int \tilde{p}(x,\varepsilon)~~ dx \right),
\end{equation}
where $\varphi(x)=\sqrt{M(x)} \psi(x)$. And we also point out that $\psi$ corresponds to the eigenfunctions of the Schr\"{o}dinger equation. Thus, the relation between two quantum momentum functions is
\begin{equation}\label{moms}
    \tilde{p}=-i\frac{M^{'}(x)}{2M(x)}+p.
\end{equation}
Introducing a general form for $ \tilde{p}$:
\begin{equation}\label{gen-p}
    \tilde{p}^{'}=a(x)+b(x)\tilde{p}+c(x)\tilde{p}^{2},
\end{equation}
which has the same form as (\ref{pdmqhj}), and using
\begin{equation}\label{mmmp}
  \tilde{p}(x)=\frac{1}{c(x)}\upsilon(x)-\frac{b(x)}{2c(x)}-\frac{c^{'}(x)}{2c^{2}(x)}
\end{equation}
in (\ref{gen-p}), we have then,
\begin{equation}\label{ric}
    \upsilon^{'}(x)-\upsilon^{2}(x)=G(x),
\end{equation}
where $G(x)$ is given by
\begin{equation}\label{G}
    G(x)=ac-\frac{b^{2}}{4}+\frac{b^{'}}{2}-\frac{b c^{'}}{2c}-\frac{3}{4}\frac{c^{'2}}{c^{2}}+\frac{c^{''}}{2c},
\end{equation}
and primes denote derivatives with respect to $x$. If we compare (\ref{ric}), (\ref{G}), (\ref{gen-p}) with (\ref{pdmqhj}), we can express
$a, b, c$ in terms of the physical quantities:
\begin{eqnarray}\label{abc}
 a(x) &=& -i M(x)(V_{eff}(x)-\varepsilon) \\ \nonumber
  b(x) &=& \frac{M^{'}(x)}{M(x)} \\ \nonumber
  c &=& -i.
\end{eqnarray}
Then, $G(x)$ can be written as
\begin{equation}\label{GG}
    G(x)=-\frac{4M^{3}(x)(V_{eff}(x)-\varepsilon)+3M^{'2}(x)-2M(x)M^{''}(x)}{4M^{2}(x)}.
\end{equation}
 \subsection{A short discussion on the multiplicities}
$G(x)$ may have the number of poles because of the mass and effective potentials. In the previous works, there were not any comments on the multiplicity of the poles \cite{bh, ranj, ozlem, asim}. In terms of the multiplicities of the poles of $G(x)$, the maximum number of distinct meromorphic solutions can be discussed.\\
\emph{Theorem }: If $G(x)$ has at least one simple pole, then (\ref{ric}) admits at most one meromorphic solution.  \\
\emph{Corollary }: Let $G(x)$ is a meromorphic function, then it has at least one pole. If (\ref{ric})  has two distinct meromorphic solutions, then all poles of $G(x)$ are of even multiplicity.\\
\emph{Proof of Theorem}: Let $\upsilon$ be a meromorphic solution of (\ref{ric}) and $x_{0}$ is a simple pole of $G(x)$. Obviously, the residue of $\upsilon$ at $x_{0}$
is $-1$. Thus, there must be a neighborhood of $x_{0}$, which is $\mathcal{D}$, such that the function
\begin{equation}\label{ix}
\xi:=(x-x_{0})e^{-\int^{x}_{x_{0}}\ \chi dx^{'}}
\end{equation}
 and the function $\chi$ are analytic in a neighborhood of $x_{0}$, and $\chi$ satisfies $\chi=\upsilon+\frac{1}{x-x_{0}}$. Henceforth, $\xi$ satisfies the well-known differential equation
\begin{equation}\label{S}
    \xi^{''}+G(x)\xi=0.
\end{equation}
Suppose that the theorem is false and (\ref{ric}) has two distinct meromorphic solutions $\upsilon_{1}, \upsilon_{2}$. Let $\xi_{1},\xi_{2}$ and $\chi_{1}$, $\chi_{2}$ satisfy (\ref{S}) and (\ref{ix}) on a neighborhood $\mathcal{S}$. If $\xi_{1}$ and $\xi_{2}$ are linearly independent, then; $\frac{d}{dx}\left(\xi_{1}\frac{d\xi_{2}}{dx}- \xi_{2}\frac{d\xi_{1}}{dx}\right)=0$. From this point of view, there is a function $f$ and a constant c such that $f:=\frac{\xi_{2}}{\xi_{1}}$, $f^{'}:=\frac{c}{\xi^{2}_{1}}$, then $x_{0}$ is a simple pole of $f$. By using (\ref{ix}), $f$ becomes analytic at $x_{0}$. And there is a contradiction. \square

Here, Theorem 6.1 in Ref. \cite{nevan} is followed to introduce the Theorem given above. In fact, previous works include even multiplicity. But, there may be also simple poles depending on the mass function. We will see this in the example of the P\"{o}schl-Teller potential, discussed below.
\subsubsection{Singular Mass with simple pole:  $x_{0}=i k \pi $, $k=0,1,2,3,..$}
We now discuss the QEM-HJ model for $M(x)=\frac{1}{\sinh x}$. We introduce
\begin{equation}\label{vpot}
    V(x)=V_{1} \coth x+ V_{2}+ V_{3} \sinh x
\end{equation}
then we obtain $G(x)$ and $V_{eff}$ as
\begin{eqnarray}\label{G1}
 - G(x) &=& V_{1} \coth x \csc h x-(\alpha(\alpha+\beta+1)+\frac{1}{4})\csc h^{2} x
+\frac{V_{2}-\varepsilon}{\sinh x}+V_{3}+\frac{\beta+1}{2}+\frac{1}{4} \\
  V_{eff}(x) &=& V_{1}\coth x+V_{2} + [ V_{3}+\frac{\beta+1}{2} + (\beta+1)\csc h^{2}x -(\alpha(\alpha+\beta+1)+\beta+1)\coth^{2}x ] \sinh x. \nonumber
\end{eqnarray}
 Thus, a Riccati equation can be obtained by the appropriate transformations (given above) as
\begin{eqnarray}\label{tr}
\upsilon^{'} - \upsilon^{2}-V_{1} \coth x \csc h x+(\alpha(\alpha+\beta+1) +\frac{1}{4})\csc h^{2} x  -\frac{V_{2}-\varepsilon}{\sinh x}-V_{3}-\frac{\beta+1}{2}-\frac{1}{4} =  0.
\end{eqnarray}
There is a simple pole coming from the coefficient of $V_{2}-\varepsilon$ in the above equation. This equation is solvable for $V_{2}=\varepsilon$ as a special case. On the other hand, one can expand $G(x)$ in a Laurent series as
\begin{equation}\label{lrrr}
    G(x)=\frac{k_{1}+k_{2}}{x^{2}}+\frac{k_{3}}{x}+...
\end{equation}
where $k_{1}=-V_{1}$, $k_{2}=(\alpha(\alpha+\beta+1)+\frac{1}{4}$ and $k_{3}=\varepsilon-V_{2}$. There is one simple pole and also even multiplicity. We use the mappings,
 $\upsilon=-\sqrt{y^{2}-1}\left(\zeta(y)-\frac{y}{2(y^{2}-1)}\right)$, $y=\cosh x$, in (\ref{tr}) and we find
\begin{eqnarray}\label{vv}
    \zeta^{'}+ \zeta^{2}-\frac{1}{2(y^{2}-1)}+\frac{3y^{2}}{4(y^{2}-1)^{2}}+\frac{1}{(y^{2}-1)}
\left(\frac{V_{1} y}{(y^{2}-1)}-
    \frac{\alpha(\alpha+\beta+1)}{(y^{2}-1)}+V_{3}+\frac{1}{4}+\frac{\beta+1}{2}\right)=0.
\end{eqnarray}
Now we have to seek solutions to the above equation. The first task is to expand $\zeta$ in Laurent series and to find the residues. Then, $\zeta$ is expanded in a Laurent series as
\begin{equation}\label{lau}
 \zeta=\frac{b_{1}}{y-1}+p_{0}+p_{1}(y-1)+...
\end{equation}
and the residue $b_{1}$ at $y=1$ can be obtained,
\begin{equation}\label{res1}
    b_{1}|_{y=1}=\frac{1}{2}\left(1\pm \sqrt{V_{1}-\alpha(\alpha+\beta+1)+1}\right).
\end{equation}
At $y=-1$, the residue is
\begin{equation}\label{res11}
    b_{1}|_{y=-1}=\frac{1}{2}\left(1\pm \sqrt{-V_{1}-\alpha(\alpha+\beta+1)+1}\right).
\end{equation}
Because we cancel the term $k_{3}$, there are two solutions for each value of the residue, as mentioned in the Theorem. But, because of the integrability condition, we will choose one of them. Now, $F^{'}(y)=Q(y) F(y)$ will be used in
\begin{equation}\label{x}
    \zeta=\frac{b_{1}|_{y=1}}{y-1}+\frac{b_{1}|_{y=-1}}{y+1}+Q(y)+h,
\end{equation}
where $h$ is a constant because of Liouville`s Theorem \cite{ablowitz}. This equation will be put in (\ref{vv}) to obtain the spectrum. The constant $h=0$ can be obtained here. In the limit of $y\rightarrow \infty$, $Q(y)$ behaves as $Q(y)\rightarrow \frac{n}{y}$ and $Q^{'}+Q^{2} \rightarrow \frac{n(n-1)}{y^{2}}$. From the coefficient of $y^{-2}$ one obtains
\begin{eqnarray}\label{bbb}
    2b_{1}|_{y=1} b_{1}|_{y=-1}+2n(b_{1}|_{y=-1}+b_{1}|_{y=1})+n(n-1)+\frac{3}{8}-
V_{3}-\alpha(\alpha+\beta+1)=0.
\end{eqnarray}
Hence, from (\ref{bbb}) we have
\begin{eqnarray}\label{ene}
    V_{3}= \gamma^{2}-(\frac{1}{2}(-1+\sqrt{-V_{1}-\alpha(\alpha+\beta+1)+1} +\sqrt{V_{1}-
    \alpha(\alpha+\beta+1)+1})-n)^{2},
\end{eqnarray}
where
\begin{eqnarray}\label{ggga}
    \gamma = -\frac{\beta+1}{2}-\frac{1}{4}+\frac{1}{4}
(-1+\sqrt{-V_{1}-\alpha(\alpha+\beta+1)+1}+\sqrt{V_{1}-\alpha(\alpha+\beta+1)+1})^{2}.
\end{eqnarray}
We note that the negative sign is used for the residues because of the integrability condition.

Finally we can give the solutions by using (\ref{res1}), (\ref{res11}), (\ref{lau}) and $\varphi=e^{i \int \tilde{p} dx}$
\begin{equation}\label{fi}
    \psi_{n}(y)=(y-1)^{a_{1}-\frac{1}{2}}~
    (y+1)^{a_{2}-\frac{1}{2}}~ P^{(2a_{1}-\frac{3}{2},  ~-2a_{2}-\frac{3}{2})}_{n}(y),
\end{equation}
and
\begin{equation}\label{fri}
    \varphi_{n}(y)=(y-1)^{a_{1}-\frac{3}{4}}~
    (y+1)^{a_{2}-\frac{3}{4}}~ P^{(2a_{1}-\frac{3}{2},  ~-2a_{2}-\frac{3}{2})}_{n}(y),
\end{equation}
where $P^{\lambda_{1},\lambda_{2}}_{n}(y)$ are Jacobi polynomials and the constants are given by
\begin{eqnarray} \label{v,w}
  a_{1} &=& \frac{1}{4}(1-2\sqrt{-V_{1}-\alpha(\alpha+\beta+1)+1}+\sqrt{V_{1}-\alpha(\alpha+\beta+1)+1}) \nonumber \\
  a_{2} &=& \frac{1}{4}(-1+2\sqrt{V_{1}-\alpha(\alpha+\beta+1)+1}+\sqrt{V_{1}-\alpha(\alpha+\beta+1)+1}).
\end{eqnarray}
 P\"{o}schl-Teller potential is generally given by \cite{sukhatme}
\begin{equation}\label{PTpot}
    U(x)=A^{2}+(B^{2}+A^{2}+A) \csc h^{2} x-B(2A+1) \coth x \csc h x.
\end{equation}
Thus, if we compare $G(x)$ in (\ref{G1}) with (\ref{PTpot}), we can obtain the constants $A$ and $B$ in terms of the constants of (\ref{vpot})
\begin{eqnarray}\label{para}
  A &=& \frac{1}{2}\left(-1+\sqrt{-V_{1}-\alpha(\alpha+\beta+1)+1}+\sqrt{V_{1}-\alpha(\alpha+\beta+1)+1}\right) \\ \nonumber
  B &=& \frac{1}{2}\left(\sqrt{V_{1}-\alpha(\alpha+\beta+1)+1}-\sqrt{-V_{1}-\alpha(\alpha+\beta+1)+1}\right).
\end{eqnarray}
\subsubsection{Exponential mass with the pole: $x_{0}=-\infty$:}
We will use effective mass $M(x)=e^{-2x}$ in this case. In \cite{bagque}, the authors introduced $V(x)$  as
\begin{equation}\label{vmrs}
    V(x) = V_{0}e^{2x}-B(2A+1)e^{x}
\end{equation}
and $G(x)$ and $V_{eff}$ have the forms
\begin{eqnarray}\label{gin}
 G(x) &=&  \varepsilon e^{-2x}+B(2A+1)e^{-x}-V_{0}+2(\beta+1)+4\alpha(\alpha+\beta+1)+1 \\ \nonumber
  V_{eff}&=& (V_{0}-2(\beta+1)-4\alpha(\alpha+\beta+1))e^{2x} -B(2A+1)e^{x}.
\end{eqnarray}
If we use the mapping $\tilde{p}=i(1+\upsilon)$, $\upsilon=y (\zeta-\frac{1}{2y})$ given in  (\ref{mmmp}) and (\ref{abc}) and $y=e^{-x}$, we have
\begin{equation}\label{qhj1}
    \zeta^{'}+\zeta^{2}+\frac{1}{4y^{2}}+\frac{1}{y^{2}}\left(\varepsilon y^{2}+B(2A+1)y-V_{0}+C\right)=0,
\end{equation}
where $C=2(\beta+1)+4\alpha(\alpha+\beta+1)-1$. Now we expand $\zeta$ in a Laurent series as
\begin{equation}\label{l2}
    \zeta=\frac{b_{1}}{y}+p_{0}+p_{1}y+...
\end{equation}
and the residue $b_{1}$ became $b_{1}=\frac{1}{2}(1\pm 2\sqrt{|V_{0}-C|})$. We now discuss the residue at $y=\infty$, on the extended complex plane. Using $y=\frac{1}{t}$ in (\ref{qhj1}) and expanding $\zeta=B_{0}+B_{1}t+B_{2}t^{2}+...$ one obtains the residue $B_{1}=-\frac{B(1+2A)}{2B_{0}}$, $B_{0}=-\sqrt{-\varepsilon}$. The coefficient of $y^{-1}$ for large $y$ gives
\begin{equation}\label{en}
    V_{0}=2(\beta+1)+4\alpha(\alpha+\beta+1)-1+(A-n)^{2},
\end{equation}
where $\varepsilon=-B^{2}$ \cite{bagque}.
The solutions can be obtained following the same way as it is discussed before, if $\zeta$ is written $\zeta=\frac{b_{1}}{y}+Q(y)+h$, and $F^{'}(y)=Q(y)F(y)$ is used, then the constant $h=-\frac{1}{2}$ can be obtained. Using (\ref{qhj1}) and the residue $b_{1}$ one obtains
\begin{equation}\label{solmorse}
    \varphi_{n}(y)=y^{A-n+1} e^{-\frac{1}{2} y} L^{2(A-n)}_{n}(y),
\end{equation}
which is consistent with \cite{bagque}.
 \section{Pseudo-Hermitian Hamiltonian connected with QEM-HJ model}
Suppose that $H$ is not a Hermitian Hamiltonian. Then, $\eta$-pseudo-Hermiticity of $H$ is equivalent to the condition given by \cite{mostafazadeh}
\begin{equation}\label{pseu}
    H^{\dag}=\eta H \eta^{-1},
\end{equation}
where $\eta$ is a linear, invertible operator. The metric operator $\eta$ is not unique for a pseudo-Hermitian operator $H$. There exists a mapping from the non-Hermitian $H$ to its Hermitian counterpart $h$, through a similarity transformation $h=\rho H \rho^{-1}$, here $H$, a differential operator act in a complex function space and the mapping function is positive definite $\rho=\sqrt{\eta}$ \cite{mostafazadeh}. A non-Hermitian, $\mathcal{PT}$ symmetric Hamiltonian model is considered by Swanson and it is given \cite{swanson} as
\begin{equation}\label{sw1}
    H=\omega(a^{\dag} a+\frac{1}{2})+\alpha a^{2}+\beta a^{\dag 2},
\end{equation}
where $a$, $a^{\dag}$ generalized creation and annihilation operators,
\begin{equation*}\label{a-ah}
    a=A(x)\frac{d}{dx}+B(x), ~~~ a^{\dag}=-A(x)\frac{d}{dx}+B(x)-\frac{dA(x)}{dx} \nonumber
\end{equation*}
and $\omega, \alpha, \beta$ are real, positive constants. In \cite{Bagchitanaka}, the more general form of Swanson Hamiltonian is introduced as
\begin{equation}\label{sw}
 H=\omega(a^{\dag} a+\frac{1}{2})+\alpha a^{2}+\beta a^{\dag 2}+\gamma a + \delta a^{\dag },
\end{equation}
which can be written in a differential operator form \cite{Bagchitanaka}
\begin{equation}\label{sH}
    H=-\bar{\omega}\frac{d}{dx}A^{2}(x)\frac{d}{dx}+b_{1}(x)\frac{d}{dx}+c_{2}(x),
\end{equation}
where $\bar{\omega}=\omega-\alpha-\beta$,
\begin{equation}\label{b1}
   b_{1}(x)= (\alpha-\beta)A(x)(2B(x)-A^{'}(x))+(\gamma-\delta)A(x),
\end{equation}
and
\begin{eqnarray}\label{c2}
    c_{2}(x) & = &(\omega+\alpha+\beta)B^{2}(x)-(\omega+2\beta) A^{'}(x)B(x)-(\omega-\alpha+\beta)A(x)B^{'}(x)\\ & &+ \beta(A(x)A^{''}(x)+A^{'2}(x))
  +(\gamma+\delta)B(x)-\delta A^{'}(x)+\frac{\omega}{2}. \nonumber
  \end{eqnarray}
Let $\Psi$ and $\mathcal{E}$ are the eigenfunctions and the eigenvalues of (\ref{sH}) i.e.,
\begin{equation}\label{heq}
   H\Psi =\mathcal{E}\Psi.
\end{equation}
If $\mathbf{p}$ is introduced by
\begin{equation}\label{frakp}
    \mathbf{p}=-i \frac{\Psi^{'}}{\Psi}
\end{equation}
and if it is used in (\ref{heq}), then it yields
 \begin{equation}\label{4}
   \mathbf{p}^{'}(x)=-i~\mathbf{p}^{2}(x)+\frac{b_{1}(x)-
    2\bar{\omega}A(x)A^{'}(x)}{\bar{\omega}A^{2}(x)} \mathbf{p}(x)-i\frac{c_{2}(x)-
    \mathcal{E}}{\bar{\omega}A^{2}(x)}.
 \end{equation}
 Comparing above equation with (\ref{gen-p}), we obtain
 \begin{eqnarray}\label{43}
   a(x) &=& -i\frac{c_{2}-\mathcal{E}}{\bar{\omega}A(x)^{2}} \\ \nonumber
   b(x) &=& \frac{b_{1}-2\bar{\omega}A(x)A^{'}(x)}{\bar{\omega}A(x)^{2}} \\ \nonumber
   c &=& -i.  \nonumber
 \end{eqnarray}
Meanwhile, it is obvious that (\ref{4}) is not any type of the QHJ equation that is discussed in Sections 1 and 2, because (\ref{sH}) is not Hermitian, and $H\Psi =\mathcal{E}\Psi$ is not a regular Sturm-Liouville type differential equation. But one can find a Hermitian counterpart of (\ref{sH}) by using a convenient metric \cite{Bagchitanaka}. But here, we can find a Riccati- type equation which corresponds to a QEM-HJ equation.  We then use:
\begin{equation}\label{23}
   \mathbf{p}(x)=i\upsilon(x)-i\frac{b_{1}(x)-2\bar{\omega}A(x)A^{'}(x)}{2\bar{\omega}A^{2}(x)}
\end{equation}
in order to transform (\ref{4}) into (\ref{ric}). Thus, $G(x)$ can be obtained as
\begin{equation}\label{rc}
    G(x)=-\frac{A^{''}(x)}{A(x)}+\frac{b^{'}_{1}(x)}{2\bar{\omega}A^{2}(x)}-\frac{c_{2}(x)-
    \mathcal{E}}{\bar{\omega}A^{2}(x)}-\frac{b^{2}_{1}(x)}{4\bar{\omega}^{2}A^{4}(x)}.
\end{equation}
We note again that, the primes in equations given above denote derivatives with respect to $x$. Hence, we can introduce $\Psi(x)$ as
\begin{equation}\label{cc}
   \Psi(x)=e^{\int ^{x} dx^{'} \frac{b_{1}(x^{'})}{2\bar{\omega}A(x^{'})^{2}}} \phi(x)
\end{equation}
where $\phi(x)=e^{-\int^{x} dx^{'} \left(\upsilon(x^{'})+\frac{A^{'}(x^{'})}{A(x^{'})}\right)}$.

On the other hand, if we follow a standard procedure(for the Schr\"{o}dinger operator) \cite{quesne, roy, Bagchitanaka}, (\ref{heq})  can be transformed into an equivalent Hermitian form by means of the similarity transformation \cite{mostafazadeh}
\begin{equation}\label{her}
    h=\rho H \rho^{-1},
\end{equation}
where the mapping function reads \cite{quesne, roy, Bagchitanaka}
\begin{equation}\label{mapping}
    \rho=\exp\left(-\frac{1}{2\bar{\omega}}\int dx \frac{b_{1}(x)}{A^{2}(x)}\right)
\end{equation}
 such that $\eta=\rho^{2}$. This transformation leads to
\begin{equation}\label{eff}
    h=-\bar{\omega}\frac{d}{dx} A^{2}(x)\frac{d}{dx}+V_{eff}(x)
\end{equation}
where \cite{Bagchitanaka}
\begin{eqnarray}\label{veff}
    V_{eff}&=& \left(\frac{(\alpha-\beta)^{2}}{\bar{\omega}}+\bar{\omega} +2(\alpha+\beta)\right)B(x)(B(x)-A^{'}(x))-(\bar{\omega}+\alpha+\beta)A(x)B(x)^{'} \nonumber
   + \frac{\alpha+\beta}{2}A(x)A(x)^{''}\\ & &+\frac{1}{4}\left(\frac{(\alpha-\beta)^{2}}{\bar{\omega}} + 2(\alpha+\beta)\right)A(x)^{'2} + \nonumber
    \left(\frac{(\alpha-\beta)(\gamma-\delta)}{\bar{\omega}}+\gamma+\delta\right)\left(B(x)-\frac{A(x)^{'}}{2}\right) \nonumber
     + \frac{(\gamma-\delta)^{2}}{4\bar{\omega}}\nonumber \\ & & +\frac{\bar{\omega}+\alpha+\beta}{2}.
   \end{eqnarray}
\subsection{P\"{o}schl-Teller and Morse potentials in Swanson model}
Now we will study the effective potentials in Swanson model. If we compare (\ref{eff}) and (\ref{pdm}) it is seen that
\begin{equation}\label{AAAAAAAAA}
    A(x)=\sqrt{\frac{1}{\bar{\omega}M(x)}}.
\end{equation}
Let us discuss two different types of $A(x)$ function. If we can give an ansatz for $A(x)$ and $B(x)$, we may be able to obtain some effective potential models. In this manner, one can obtain the residues and all solutions in terms of the Swanson parameters $\omega,\alpha,\beta$.\\
\textbf{i)} Firstly, we take $\gamma=\delta=0$ for convenience. If we choose
\begin{equation*}\nonumber
    A(x)=\sqrt{\frac{\sinh x}{\bar{\omega}}}, ~~ B(x)=A(x)^{-1}
\end{equation*}
then, using (\ref{rc}), we obtain
\begin{eqnarray}\label{55}
    G(x)&=-\frac{\alpha^{2}+(\beta-2\omega)^{2}-2\alpha(\beta+2\omega)}{16\bar{\omega}^{2}}-\frac{\omega-2\mathcal{E}}{2\bar{\omega}}\csc h x+\\
    &\frac{\omega(\alpha+\beta)-4\alpha\beta}{2\bar{\omega}^{2}}\coth x \csc h x-\left(\frac{(\alpha-\beta)^{2}}{16\bar{\omega}^{2}}+\omega^{2}-4\alpha\beta-\frac{1}{4}\right)\csc h^{2} x. \nonumber
    \end{eqnarray}
and from (\ref{veff}) we have
\begin{eqnarray}\label{vefec1}
   V_{eff}(x)&=-\frac{\omega(\alpha+\beta)+4\alpha\beta}{2\bar{\omega}} \coth x+\frac{(\alpha-\beta)^{2}}{16\bar{\omega^{2}}} \coth x \csc hx+
(\omega^{2}-4\alpha\beta) \csc hx \\+& \frac{\alpha+\beta}{4\bar{\omega}}\sinh x. \nonumber
\end{eqnarray}
The QEM-HJ equation is solvable for $\omega=2\mathcal{E}$. Comparing (\ref{55}), (\ref{ric}) and (\ref{PTpot}) we get
\begin{eqnarray}\label{ABBA}
  A &=& -\frac{(\alpha+\beta)(\alpha+\beta-3\omega)+\omega^{2}+4\alpha\beta}{2\alpha(\alpha+\beta-\omega)^{2}} \\ \nonumber
  B &=& -\frac{(\alpha+\beta)^{2}-\omega(\alpha+\beta)(\alpha+3)+\omega^{2}+4(\alpha-1)\alpha\beta}
  {2\alpha(\alpha+\beta-\omega)^{2}}\nonumber.
\end{eqnarray}
$V_{1}$, $V_{2}$ and $V_{3}$, the parameters of (\ref{vmrs}), can also be written in terms of Swanson parameters as
\begin{eqnarray}
  V_{1} &=& -\frac{\omega(\alpha+\beta)-4\alpha\beta}{2\bar{\omega}} \\ \nonumber
  V_{2} &=& \omega \\ \nonumber
  V_{3} &=& \frac{\alpha^{2}+(\beta-2\omega)^{2}-2\alpha(\beta+2\omega)}{16\bar{\omega}^{2}}
  +\frac{1}{4}+\frac{\tilde{\beta}+1}{2}\\
  \varepsilon&=&2\mathcal{E}, \nonumber
\end{eqnarray}
where $\tilde{\beta}$ is used instead of $\beta$ that is the parameter given in  von Roos Hamiltonian. The solutions $\Psi$ which is defined in (\ref{cc}) is then given by
\begin{equation}\label{fi}
    \Psi_{n}(x)\sim
     (\cosh x-1)^{\frac{B-A}{2}-\frac{(\alpha-\beta)(1-4\bar{\omega})}{8\bar{\omega}}} (\cosh x+1)^{-\frac{B+A}{2}-\frac{(\alpha-\beta)(1+4\bar{\omega})}{8\bar{\omega}}}P^{(B-A-\frac{1}{2},-B-A-\frac{1}{2})}_{n}(\cosh x),
\end{equation}
where the constants $A$ and $B$ were given by (\ref{ABBA}) and $0\leq x\leq \infty$. In this case we note that the metric is given by
\begin{equation}\label{metr}
    \eta= (\sinh x)^{\frac{\alpha-\beta}{\bar{\omega}}} \left(\tanh \frac{x}{2}\right)^{-2(\alpha-\beta)}
\end{equation}
such that the Hermitian inner product  $\ll\Psi(x)|\Psi(x)\gg_{\eta}=<\Psi(x)|\eta|\Psi(x)> $ can be used for the normalization.
The constraints for the parameters are given by $B-A> \frac{(\alpha-\beta)(1-4\bar{\omega})}{8\bar{\omega}}$ which is satisfied by the relations $A$ and $B$ in (\ref{ABBA}).\\
\textbf{ii)} Next we introduce $A(x)$ and $B(x)$ as
\begin{equation*}\label{ab}
    A(x):= \frac{e^{x}}{\sqrt{\bar{\omega}}} =:B(x).
\end{equation*}
In this case we obtain $G(x)$ as
\begin{equation}\label{21}
    G(x)=-\frac{(\alpha-\beta)^{2} }{4\bar{\omega}^{2}}+\frac{2(\beta\gamma+\alpha\delta)
-(\delta+\gamma)\omega}{2\bar{\omega}^{3/2}}e^{-x}+
    \left(-\mathcal{E}+\frac{\omega}{2}-\frac{(\gamma-\delta)^{2}}{4\bar{\omega}}\right)e^{-2x}
\end{equation}
and the effective potential is given by
\begin{equation}\label{morseEff}
    V_{eff}=\frac{(\gamma-\delta)^{2}+2\omega \bar{\omega}}{4\bar{\omega}}
    +\frac{\left(\frac{(\alpha-\beta)(\gamma-\delta)}{\bar{\omega}}\right)+\gamma+\delta}{2\sqrt{\bar{\omega}}} ~ e^{x}+\left(-1+\frac{(\alpha-\beta)^{2}}{4\bar{\omega}^{2}}\right)e^{2x}.
\end{equation}
By comparing (\ref{21}) and (\ref{gin}) we obtain
\begin{eqnarray}
  A &=&  -\frac{1}{2}+\frac{2(\beta\gamma+\alpha\delta)-(\delta+\gamma)}{4\bar{\omega}^{\frac{3}{2}}} ~B^{-1}  \\ \nonumber
  B &=& \left( \mathcal{E}-\frac{\omega}{2}+\frac{(\gamma-\delta)^{2}}{4\bar{\omega}}\right)^{\frac{1}{2}} \\
  \varepsilon &=& \mathcal{E} \nonumber
\end{eqnarray}
and
\begin{equation}\label{v00}
    V_{0}=2(\tilde{\beta}+1)+4\tilde{\alpha}(\tilde{\alpha}+\tilde{\beta}+1)-1+\left(\frac{\alpha-\beta}{2\bar{\omega}}\right)^{2}, ~~~~\frac{\alpha-\beta}{2\bar{\omega}}=A-n,
\end{equation}
where $\tilde{\alpha}$, $\tilde{\beta}$ are the constants defined in the von Roos Hamiltonian instead of $\alpha$, $\beta$.
And the complete solutions are given by
\begin{equation}\label{09}
    \Psi_{n}\sim (e^{-x})^{A-n-\frac{\alpha-\beta}{2\bar{\omega}}} e^{(-B+\frac{\delta-\gamma}{2\sqrt{\bar{\omega}}}) e^{-x}} L^{2(A-n)}_{n}(2B e^{-x}).
\end{equation}
In this case, $\eta$ can be given as
\begin{equation}\label{et2}
    \eta=e ^{-\frac{\alpha-\beta}{\bar{\omega}} x+ \frac{\gamma-\delta}{\sqrt{\bar{\omega}}} e^{-x}}
\end{equation}
which can be used in the Hermitian inner product $\ll\Psi(x)|\Psi(x)\gg_{\eta}=<\Psi(x)|\eta|\Psi(x)> $ to obtain normalized wave functions. We note that the boundary conditions imply $A>n$ and $-\infty \leq x \leq \infty$.
\section{Conclusions}
In this work we have studied the generalized QHJ equation within the context of effective mass approach. We have used $\tilde{p}$ instead of $p$ as an effective mass-quantum momentum function which can also be used to obtain the solutions of a position dependent Schr\"{o}dinger equation, without solving it. We have obtained the solutions of the QEM-HJ equation for two different types of mass function. It is seen that if there is an even multiplicity of the poles, there are at most two solutions but only one of them is acceptable by reason of the physical conditions.  On the other hand, the mass dependent QHJ equation differs from the regular QHJ equation because of the pole contribution of the mass term. This can be observed in the P\"{o}schl-Teller example such that the coefficient of the single pole, $k_{3}$, is taken as zero. But, in the example of the exponential mass, the pole structure of $G(x)$ does not lead to such a restriction.

We have emphasized that the transformation in (\ref{mmmp}) that maps the QEM-HJ equation into a normal Riccati equation is also used for the Swanson model. This also means that an eigenvalue equation of the non-Hermitian system can be mapped into a Hermitian one by (\ref{mmmp}) which is seen to be in Riccati equation form. We have also shown that, using certain choices of $A(x)$ and $B(x)$, leads to some effective potentials generated from the Hermitian counterpart of the Swanson Hamiltonian. The solutions of (\ref{4}) are obtained for the P\"{o}schl-Teller and Morse potential cases without solving the QEM-HJ equation. To obtain the normalized wave functions, we use  the $\eta$- Hermiticity condition for (\ref{fi}) and (\ref{09}). On the other hand, it is clear that an extensive class of soluble potentials can be classified according to choices of $A(x)$, $B(x)$. It would be an interesting problem to discuss a relativistic quantum stationary Hamilton- Jacobi equation within effective mass approach.

\section{Acknowledge}
The author wishes to thank Dr. Richard L. Hall for reading this manuscript carefully. The author also thanks Dr. Pinaki Roy for bringing the topic to her attention.
\\

\end{document}